\documentclass[12pt, a4paper, twocolumn]{article}
\usepackage{graphicx}
\usepackage{parskip}
\usepackage{caption}
\usepackage{hyperref} 
\usepackage{float}
\usepackage{amsmath}
\usepackage{mathtools}
\usepackage{latexsym}
\usepackage{lipsum}
\usepackage{abstract}
\usepackage{sectsty}
\sectionfont{\fontsize{15}{15}\selectfont}
\subsectionfont{\fontsize{12}{15}\selectfont}
\usepackage[backend=biber, style=numeric, sorting=nty, maxnames=20]{biblatex}
\usepackage[top=1in]{geometry}
\DeclareCaptionFormat{myformat}{
    \fontsize{10}{11}\selectfont#1#2#3}
\title{Security Survey and Analysis of Vote-by-Mail Systems}
\date{} 

\author{
\scalebox{0.9}{\begin{tabular}{c} Jenny Blessing \\ {\tt jbless@mit.edu} \end{tabular}} \and
\scalebox{0.9}{\begin{tabular}{c} Julian Gomez \\ {\tt jrgomez@mit.edu}\end{tabular}} \and
\scalebox{0.9}{\begin{tabular}{c} McCoy Pati\~{n}o \\ {\tt mccoyp@mit.edu} \end{tabular}} \and
\scalebox{0.9}{\begin{tabular}{c} Tran Nguyen \\ {\tt kiretran@mit.edu} \end{tabular}} 
}

\begin{document}
\vspace{-11cm}
\maketitle
\vspace{-15em}

\section*{Abstract}
\vspace{-1.5em}
Voting by mail has been gaining traction for decades in the United States and has emerged as the preferred voting method during the COVID-19 pandemic \cite{VBM, guardian}.  In this paper, we examine the security of electronic systems used in the process of voting by mail, including online voter registration and online ballot tracking systems.  The goals of these systems, to facilitate voter registration and increase public confidence in elections, are laudable.  They indisputably provide a critical public good.  It is for these reasons that understanding the security and privacy posture of the mail-in voting process is paramount.

\vspace{-.2em}
We find that online voter registration systems in some states have vulnerabilities that allow adversaries to alter or effectively prevent a voter's registration.  We additionally find that ballot tracking systems raise serious privacy questions surrounding ease of access to voter data.  While the vulnerabilities discussed here are unlikely to enable an adversary to modify votes, several could have the effect of disenfranchising voters and reducing voter confidence in U.S. elections infrastructure, thereby undermining the very purpose of these systems.

\section{Introduction}
\vspace{-1.6em}
In an era where COVID-19 has necessitated social distancing and an elimination of large gatherings, the logistics of political elections in the United States are a natural cause for concern.  On the one hand, it is important that our democratic processes proceed as normal and elections continue to take place; on the other, in-person voting at centralized locations poses a potential health threat to citizens and threatens to suppress voter turnout.

This leaves two possibilities for remote voting: Internet voting or voting by mail.  Voting over the Internet has repeatedly been shown to be dangerously insecure by security researchers, leaving large-scale mail-based voting as the only viable remote option \cite{internetVoting, nationalAcademies, voatz}.  Voting by mail allows citizens to exercise their right to vote from the safety of quarantine.  There are currently five states---Colorado, Oregon, Washington, Hawaii, and Utah---that conduct elections almost entirely by mail, and an additional six provide a permanent mail ballot option \cite{scaleVAH}.  While only a handful of states currently vote primarily by mail, U.S. Senators Amy Klobuchar and Ron Wyden introduced a bill in March 2020 that would ``guarantee every voter a secure mail-in paper ballot" \cite{WydenKlobuchar, scaleVAH}.

To reassure voters that their mailed ballot is on its way or that their returned ballot was counted, states that make heavy use of voting by mail have widely adopted online ballot tracking systems \cite{scaleVAH}.  These systems generally allow a voter to track the status and location of his or her ballot at any point and receive notifications by email or SMS.

The novelty of these tracking systems is such that none of them have yet been publicly evaluated from a technical perspective.  In this paper, we hope to bridge this gap and provide an evaluation useful for private citizens concerned about their privacy as well as for election administrators interested in ensuring the integrity of their elections.  The ongoing pandemic serves as a reminder that the security of remote systems that support the voting process, such as online voter registration systems and ballot tracking systems, is equally as important as the security of in-person vote-casting systems.

\begin{figure*}[!ht]
    \label{BallotTRACE}
    \centering
    \includegraphics[width=\textwidth]{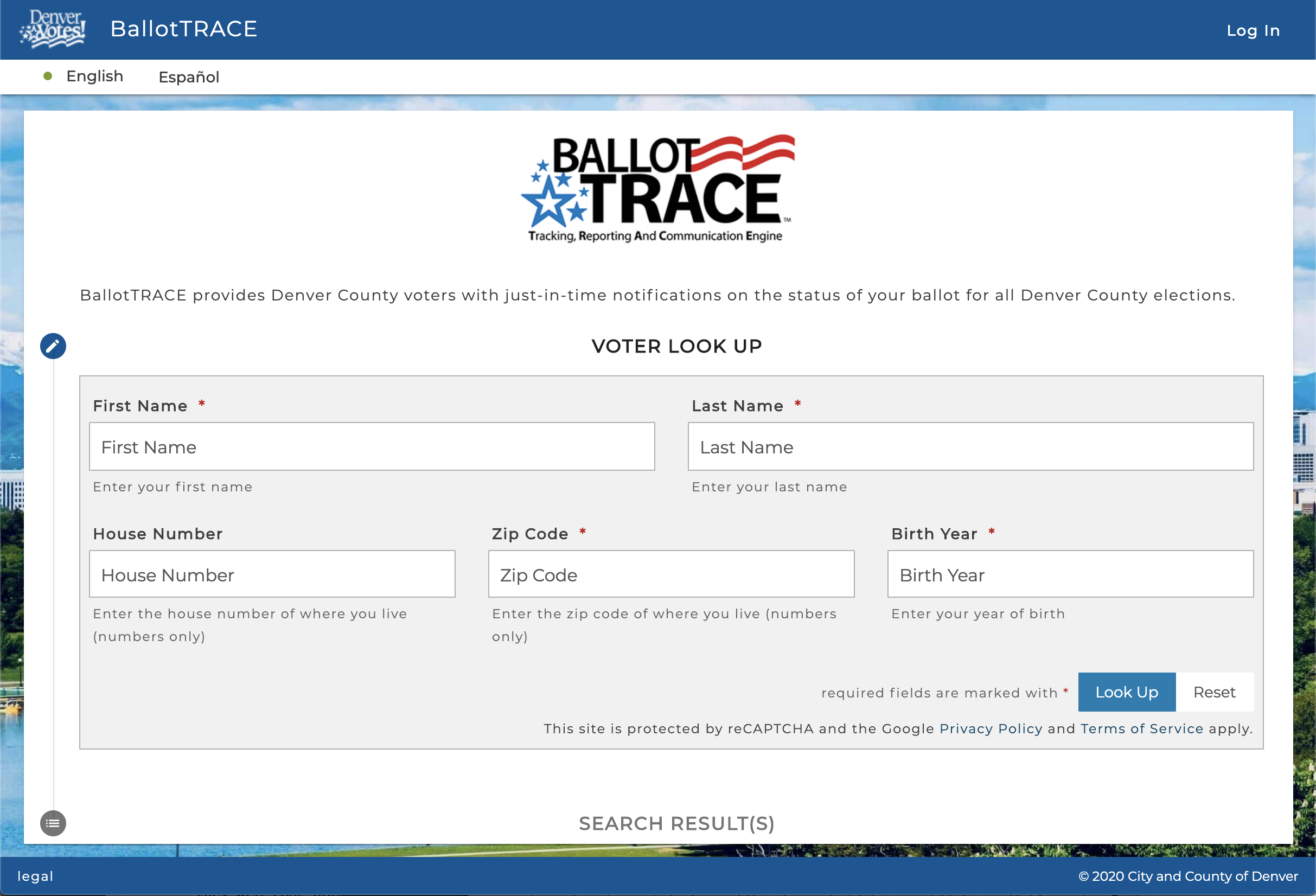}\\
    \caption{BallotTRACE's voter lookup search form.}
\end{figure*}

\subsection{Existing Ballot Tracking Systems}
While only five states conduct elections entirely by mail, all states allow absentee balloting pending an acceptable excuse, and so ballot tracking systems are used across several states and counties.  There are three ballot tracking systems currently in common use: Ballot Scout, BallotTRACE, and BallotTrax.  By our count, around 15 states have counties that use at least one of these three tracking websites.

Two of these three systems, BallotTRACE and BallotTrax, were originally developed in Colorado's Denver County, which has been the vanguard of voting by mail.  Each of these systems use USPS Intelligent Mail barcodes (IMb) to track a ballot from when it is mailed from a centralized election facility to when the completed ballot is received by local voting officials. The services and tracking capabilities offered by all three systems are effectively the same. As an example, the web interface of BallotTRACE is shown in Figure 1.

\textbf{Ballot Scout}  Ballot Scout is a web application developed by Democracy Works, a nonpartisan, non-profit organization that provides tools for voters and support to election officials \cite{democracyWorks}.

\textbf{BallotTRACE}  BallotTRACE is a web application developed by the Denver Elections Division in 2009 in partnership with a local software company, i3logix \cite{ballotTRACEProgram}. It was the first of the three systems.

\textbf{BallotTrax}  BallotTrax is available as a web application and as an iOS mobile application.  It is a spin-off of BallotTRACE, marketed more widely by i3logix.  It is run as a for-profit service \cite{fastCompany}.

\section{Online Voter Registration}

\subsection{Current State of OVR}

A voter must first submit a voter registration form in all states but North Dakota in order to mail in a ballot. This is traditionally done in person or by mail, but starting with Arizona in 2004, voter registration has increasingly moved online. As of February 3, 2020, 39 states and the District of Columbia allow online voter registration (OVR) in some form. Cost savings associated with OVR are often cited by election officials as a significant reason for the shift, but other perks include task automation and greater convenience for voters \cite{pewTrustsUnderstanding, rockTheVote}. Verifying identities with personally identifying information (PII) over the internet and protecting this information, however, requires some careful consideration.

In many aspects, OVR mirrors mail-in registration. Voters enter their name, date of birth, and some PII that only the voter is presumed to know. It is most common for states to require only a driver's license/permit or state ID number for this PII \cite{pewTrustsOVR}. This would be cause for concern when an individual's name and date of birth are used to generate these numbers, as is the case in 11 states \cite{uniqueID}. Maryland is one of these state, but allows voters to register using only this ID number---the last four digits of a voter's Social Security number (SSN) are only required if they do not have a state ID \cite{mdRegistration}. This clearly presents a potential avenue for voter impersonation.

Many states require additional PII to make their systems more secure. This usually means requiring a voter's SSN or its last four digits, but some states opt for an audit code or ID issuance date instead. Unfortunately, none of this PII is entirely secure.

\subsection{Security Concerns}

Sweeney et al. evaluated how voter identity theft done with OVR. They found that much of the data required for malicious registration is publicly available or can be obtained via data brokers and the dark web. They found that 1\% of nationwide registrations could be targeted with data costing only \$10,081 to \$24,926 in total, depending on the source \cite{harvardPaper}. Disenfranchising 1\% of voters could feasibly affect the outcome of national elections if targeted correctly.

The authors acknowledge that the security risks with online registration are not particularly new, but that the digital process makes it easier to carry out such attacks on a larger scale. In order to prevent large-scale identity theft attacks through automation, the National Conference of State Legislatures (NCSL) recommends using CAPTCHA on registration websites \cite{ncslSecuring}.

CAPTCHA provides some defense against automated registration, but the researchers point out that this defense is being weakened by the advancement of machine vision algorithms. Programs developed by Google, academic researchers, and other companies can bypass a variety of CAPTCHAs and re-CAPTCHAs with 90+\% accuracy, making them only a ``nominal deterrent" \cite{harvardPaper}.

It would appear, then, that securing voter registration requires more than securing just registration forms and websites. This is highlighted by a reported incident during the 2016 presidential primary election in Riverside County, California. District Attorney Michael Hestrin ordered an investigation when 20 formal complaints were received on election day, with voters claiming that they were turned away from the polls due to changes in their party registration that they had not made. The investigation found that registrations were altered through California's registration website, but no IP addresses were collected and no audits, if any were performed, revealed suspicious activity \cite{harvardPaper, primaryHack}.

\vspace{-.5em}
\subsection{Security Recommendations}

Audits should be routinely performed on voter registration records in order to detect an unusual volume of activity, as the NCSL recommends \cite{ncslSecuring}. We also recommend recording IP addresses that are used when making registration changes so investigations can make headway if they are necessary. Providing confirmation of registration changes by any available means of contact could also alert voters to suspicious activity. For instance, an example confirmation would be to send a notice to a voter's old and new addresses when address changes are made online \cite{ncslInterview}.

When transferring data through online registration forms, sensitive PII should be end-to-end encrypted to minimize the risk of adversaries capturing this data and using it to modify registrations. The National Institute of Standards and Technology (NIST) has specified best security practices when handling election materials, including registration data. NIST recommends that states use TLS 1.0 or above to encrypt transmitted registration data \cite{nist}. We would update this recommendation to suggest using TLS 1.2 or above, given that most browser support for 1.0 and 1.1 will be dropped soon due to security vulnerabilities in each.

\subsection{OVR Encryption Evaluation}

We used an online server testing tool provided by Qualys, Inc. to evaluate the encryption protocol security of the OVR websites provided by each state and the District of Columbia \cite{sslServerTest}. With the exception of Alaska's website, each website received a ``B" rating or higher from the tool. Alaska's website received an ``F" for its vulnerability to Zombie POODLE attacks that allow some plaintext reading and encrypted block reorganization \cite{zombiePOODLE}.

Two states' websites---Florida's and Pennsylvania's---demonstrate a vulnerability in their use of Diffie-Hellman key exchange that allows a man-in-the-middle attack known as ``Logjam". This Logjam attack allows an adversary to read and modify data passed over the connection \cite{weakDH}.

Five states' websites---Iowa's, Kentucky's, Nebraska's, New York's, and Pennsylvania's---don't use forward secrecy. Without forward secrecy, an adversary who discovers a server's private key can use it to decrypt any and all past messages sent over the channel \cite{forwardSecrecy}.

Unfortunately, 14 websites support TLS 1.0 and 20 support TLS 1.1. One website---West Virginia's---also provides undesirable support for SSL 3. The good news is that all websites support TLS 1.2, and 10 even provide support for TLS 1.3. 

\section{USPS Services}

The United States Postal Service (USPS) is the infrastructural backbone that provides chain of custody service for ballots and related election mail. It has two main services utilizing the Intelligent Mail barcode (IMb): Informed Delivery (ID) and Informed Visibility - Mail Tracking \& Reporting (IV-MTR). Informed Delivery is the older and original initiative by the Postal Service to improve transportation transparency, while Informed Visibility is a service and corresponding API provided to business owners. Both attempt to provide end-to-end tracking, with a few differences in implementation.

Should Informed Delivery be compromised, its utility in performing “wholesale fraud” is at best, negligible. We primarily assess ID and IV-MTR as a model to inform us on the availability and accessibility of ballot-tracking services, as well as its accuracy and confidentiality measures. Security weaknesses in IV-MTR pose a slightly larger threat, but do not point to dire security dilemmas in using vote-by-mail. As imperative as the USPS is to scaling up vote-by-mail, we feel it worthy to discuss past security oversights in these systems and what has been addressed since \cite{scaleVAH}. Security concerns regarding the Intelligent Mail barcode will be discussed in \S4.2, Barcode Security; this section will focus on the ID and IV-MTR services explicitly provided by USPS.

\subsection{Informed Delivery}

Informed Delivery was originally piloted in 2014 for a few select zip codes, and as of 2017 provides customers in most major zip codes with the ability to determine where their mail is in shipment. Information provided through Informed Delivery includes location information based on scans of the parcel’s barcode at each transfer point, and a grayscale image of the front of the parcel. The need to scan each individual parcel results in poor real-time performance, with users of Informed Delivery noting that the delivery estimates are often not reliable, or mail updates coming in much later than expected. Performance optimizations were made to address these issues, making Informed Visibility a more performant ``real-time tracker” by not requiring finer-granularity barcode tracking. The official site for accessing Informed Visibility states it ``leverages intelligence to create logical and assumed handling events to provide expanded visibility,” or makes reasonable assumptions regarding a parcel’s location based on the movements of its expected carrier, with any additional confirmation provided by scanning the parcel itself \cite{usps2019APIcopy, uspsIVsite}.

Prior to early 2019, USPS Informed Delivery did not rigorously authenticate identity before allowing users to access the tracking service. Account creation used a knowledge-based authentication (KBA) scheme, using approximately 4 multiple-guess questions using information from credit-bureaus. This security scheme was woefully lacking and led to a prolific string of stalking, credit card fraud, and identity theft cases in 2017-2018 \cite{krebsID2018}. USPS hesitated to implement proposed security schemes, including utilizing its own postage service to mitigate the widespread attacks for nearly two years \cite{krebsID2017}. USPS strongly urged users proactively make accounts with strong passwords to counter the fraudsters, and closing fraudulent accounts required users to send sensitive security question information to customer support through email \cite{krebsID2018}.

As of early 2019, we find that much of the earlier concerns have largely been addressed after several iterative failures in addressing the weak security. Accessing Informed Delivery no longer relies solely on KBA; services associated with Informed Delivery are now decoupled from general account privileges and require an extra one-time two-factor authentication to access them. Per the January 2020 USPS Informed Delivery sign-up guide, Informed Delivery is not available for businesses, while personal use requires a valid address or P.O. box in an eligible location \cite{uspsSignup}. Eligible locations allow three possible avenues for registering for Informed Delivery. 

Two of the avenues are given online, with the third in fine print. Upon attempting to view tracking information, a user is prompted with the two main signup options after account creation: a one-time code sent to phone via SMS or to request for a code to be mailed to the registered address. The more convenient method utilizes mobile account information from carriers including AT\&T, T-Mobile, Verizon, U.S. Cellular, and other branded wireless operators within the United States. USPS account profile information must match with regards to address, name, and number as information provided by the carrier before a one-time passcode is sent \cite{uspsMobilePolicy}. For this modality of verification, an attack would require the account password, as well as a phone number associated with the correct address. There are no limits or checks placed on changing Account Profile information after logging in, but a temporary lockout is placed on attempting to verify by phone after 3 changes in a day. Should an account be verified, upon changing, prior verification is nullified and re-authentication must be done. 

\begin{figure*}
    \label{mail} 
    \centering
    \includegraphics[width=\textwidth]{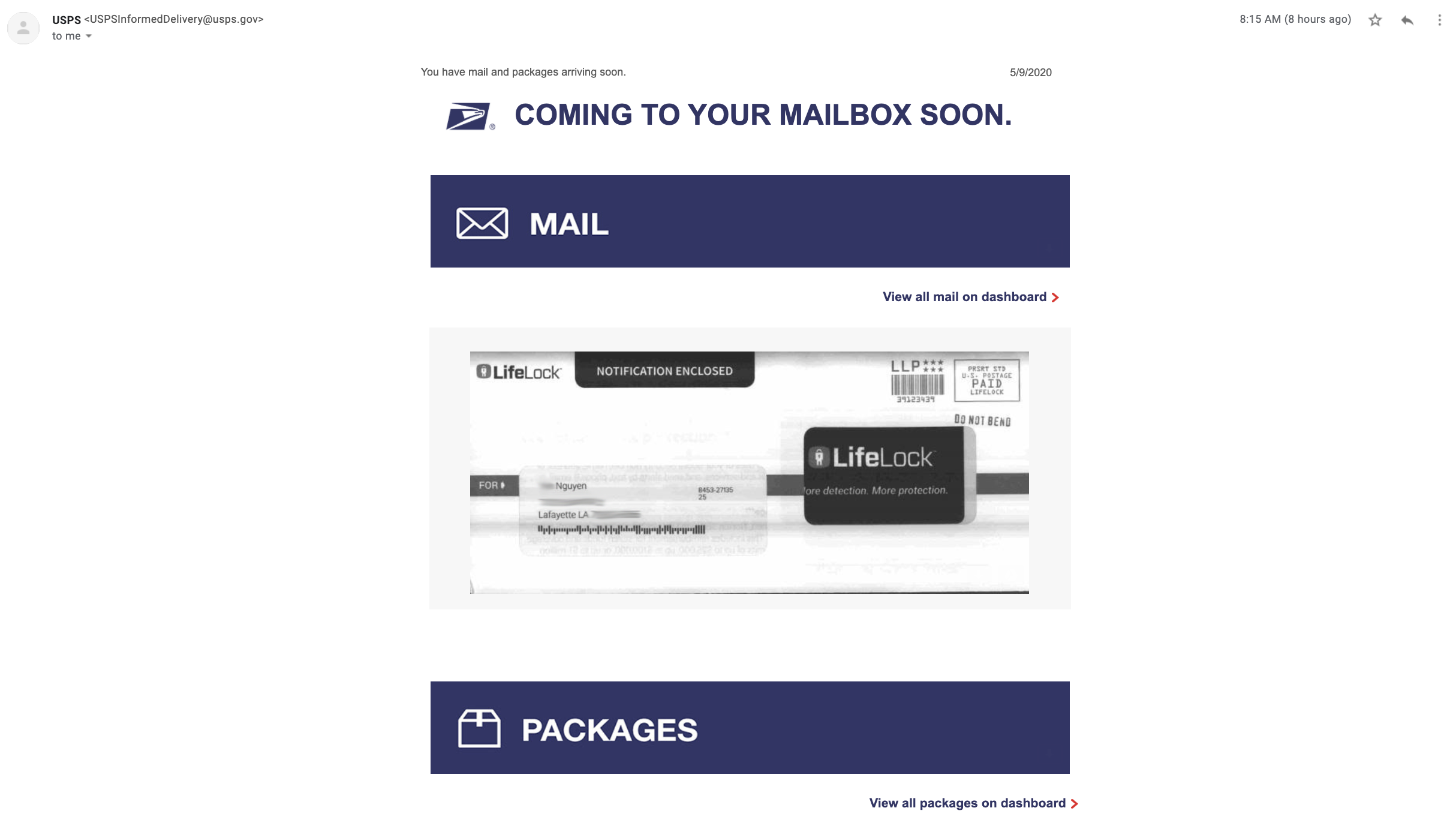}\\
    \caption{Email received from using one member's phone to sign-up at another member's address. Grayscale images of mail and incoming parcel bundles are shown. Sensitive information redacted.}
\end{figure*}

From testing with a toy account, information is simply checked against the service/billing name and address associated with the phone number for authentication. A motivated attacker could likely change their address through their service provider as we did using a volunteer's Google Fi account to switch addresses and sign up for one of our parent's residences. A screenshot from the email associated with this toy account of a recent Informed Delivery email is shown below in \hyperref{mail}[Figure 1]. The grayscale image has been partially censored to hide sensitive information for the purposes of this report and is unaltered in the email. The email service is an opt-out feature of Informed Delivery. 

The second option presented is to request for an invitation code to be sent to the specified address by mail. For those unable to do either, USPS also states in fine print that they can authorize accounts by walk-in at participating locations with proper identification.

To briefly cover accessibility with ID, apartment addresses within eligible zip codes are frequently ineligible for Informed Delivery sign-up. The third authentication option, of allowing users to authenticate their identities at select postal service locations is likely not viable for certain zip codes, and are entirely unavailable at the time of this writing. Vote-by-mail access has been a concern for populations without a permanent address or P.O. box. Voter registration and access to tracking or registration services that require permanent addresses or a particular locality have seen poor registration and disproportionately poorer turnout rates among minority populations \cite{aclu}. North Dakota, a prominent and controversial example of voting inaccessibility, agreed to a court-order to ease registration restrictions on the basis of address after nearly four years of litigation and only a week prior to the writing of this section \cite{NARF}.

\subsection{Informed Visibility}

Whereas security issues generally stem from user authentication for Informed Delivery, Informed Visibility-Mail Tracking \& Reporting’s security concerns stem from its API. Ballot tracking applications we examined do not explicitly state whether they use IV-MTR’s API to inform their mail-status updates and announcements, so we will hit main points of concern with IV-MTR. 

IV-MTR returns multiple file formats, from PKG to JSON, containing parcel location information. Although we were unable to access the current API documentation, an older copy of the API \cite{usps2018APIcopy} and a partially retracted 2018 security audit on IV-MTR implies numerous security and encryption weaknesses, system misconfigurations on each of the 13 IV-servers, among other concerns \cite{uspsAudit, usps2018APIcopy}. Later the same year, news stories reported that the API accepted wildcard search parameters for nearly every method and did not authenticate a queries’ viewing permissions before returning relevant data \cite{krebsIVMTR}. 

Poignantly, a report querying ``for readers who volunteered to help with this research" was able to gain access to ``multiple accounts when those users had more than one user signed up at the same physical address" \cite{krebsIVMTR}. The security audit only occurred a few weeks before the news break, and the allowance of unverified wildcard search queries is a non-trivial oversight. Assuming security vulnerabilities have been patched, the information a ballot tracing app has access to through these APIs is not clear; whether the app stores non-election related parcel information is also of concern.

A scan of the 2019 copy of the IV-MTR documentation published after the security patch shows that although connection is still only secured with TLS 1.0, the authentication protocol now requests user information in search queries, and an authentication token time-out after 15 minutes.  \cite{usps2019APIcopy}. 

\section{Ballot Tracking Systems}
\subsection{Tracking System Authentication}
All three major ballot tracking systems---Ballot Scout, BallotTRACE, and BallotTrax---have online web applications that allow a voter to view their ballot tracking status \cite{caliBallotTrax, ballotTRACE, ballotScoutLookup}.  These lookup systems authenticate users using only voter record data that is publicly available in many states, however, enabling users other than the voter in question to view the voter’s ballot status and, perhaps of greater concern, voting history.

States that make voter files public have historically done so to allow public scrutiny to prevent voter fraud, but political campaigns have also benefited greatly from the availability of voter databases \cite{voteShaming, stateShaming}.  Others have quickly capitalized on this data.  In 2018, two mobile applications, VoteWithMe and OutVote, were released.  These services used information from government records to allow consumers to see whom of their friends and family voted in recent elections by matching the smartphone's contacts to voter files \cite{voteShaming}, with the effective end goal of using social pressure to get people to vote.  While these apps have lost popularity since the November 2018 election, ballot tracking websites provide very similar information and have renewed this conversation.

\begin{figure*}
    \label{mayor_info} 
    \centering
    \includegraphics[width=\textwidth]{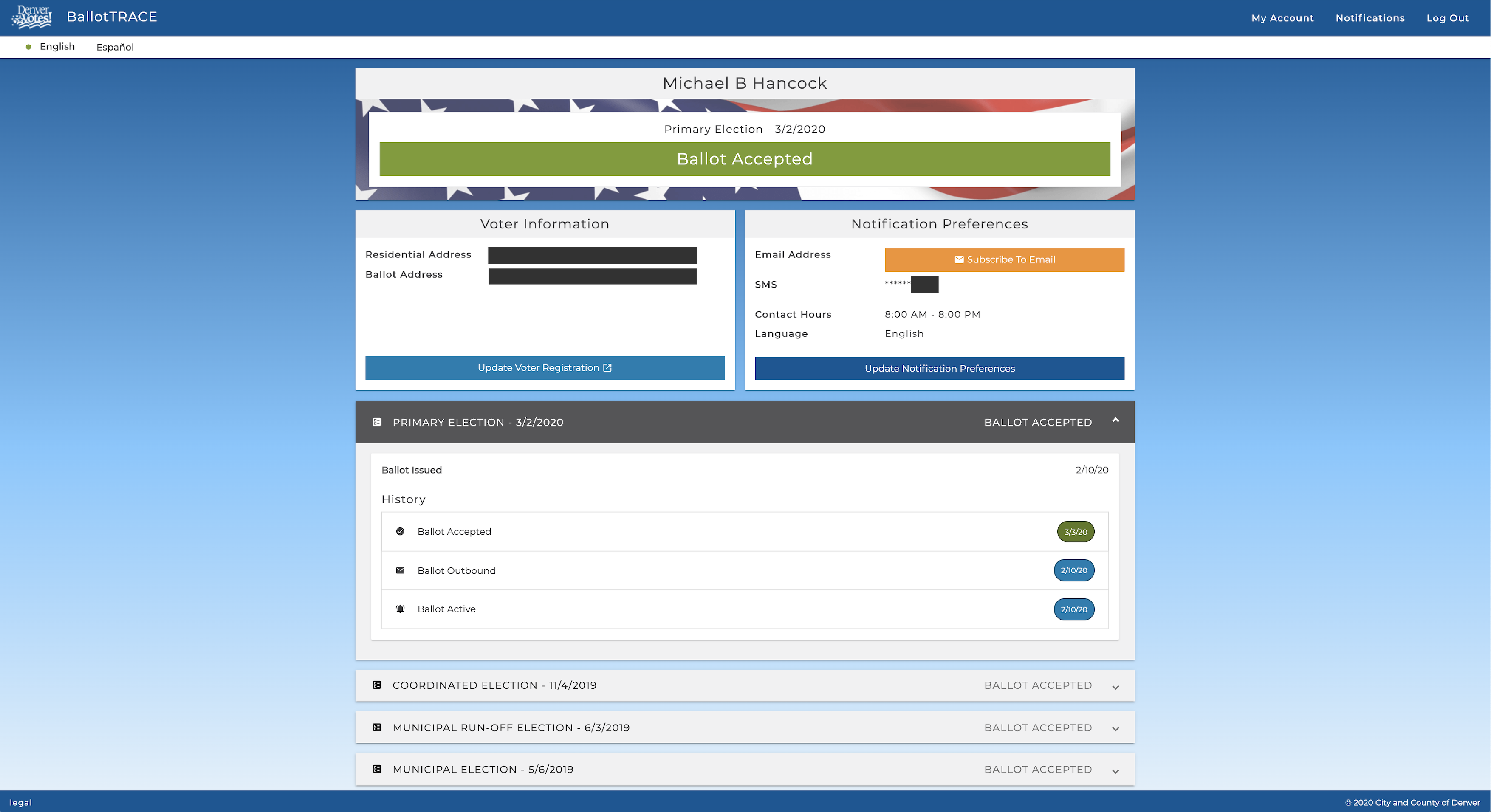}\\
    \caption{BallotTRACE's lookup page for the mayor of Denver, CO, that displays partial voting history indicating that the mayor cast a ballot in the 2020 primary elections, accessed using public voter records. Sensitive information is redacted.}
\end{figure*}

For example, California’s version of BallotTrax advertises on its homepage that, when using its system, ``tracking your ballot...has never been easier" \cite{caliBallotTrax}.  Unfortunately, this ease comes with a security tradeoff.  BallotTrax asks voters for their first name, last name, date of birth, and ZIP code in order to view tracking information, all data contained in California’s voter database which has been made publicly available under freedom of information requests \cite{caliBallotTrax}.

Sites such as VoterRecords.com have taken voter databases from several states and collected the information under one centralized website, allowing any user to search for a voter’s record using only their name, to view all voters registered in a particular district, and various other combinations.  According to its website, VoterRecords.com is ``sourced from official government public records that were released under FOIA and public record laws" \cite{voterRecords}.

Although this site has collected the records of just 16 out of 50 states, numerous other states have their voter databases separately available online.  New York, for example, is not included in VoterRecords.com, but in 2019 the New York City Board of Elections uploaded voter enrollment data for 4.6 million voters to its website \cite{nycVoter}.  While it appears that this particular database has since been removed, the state allows anyone to request the full voter list, and private citizens have made a handful of New York voter lookup tools available online \cite{nycDatabase}.

\begin{figure*}[!ht]
    \label{notification} 
    \centering
    \includegraphics[width=\textwidth]{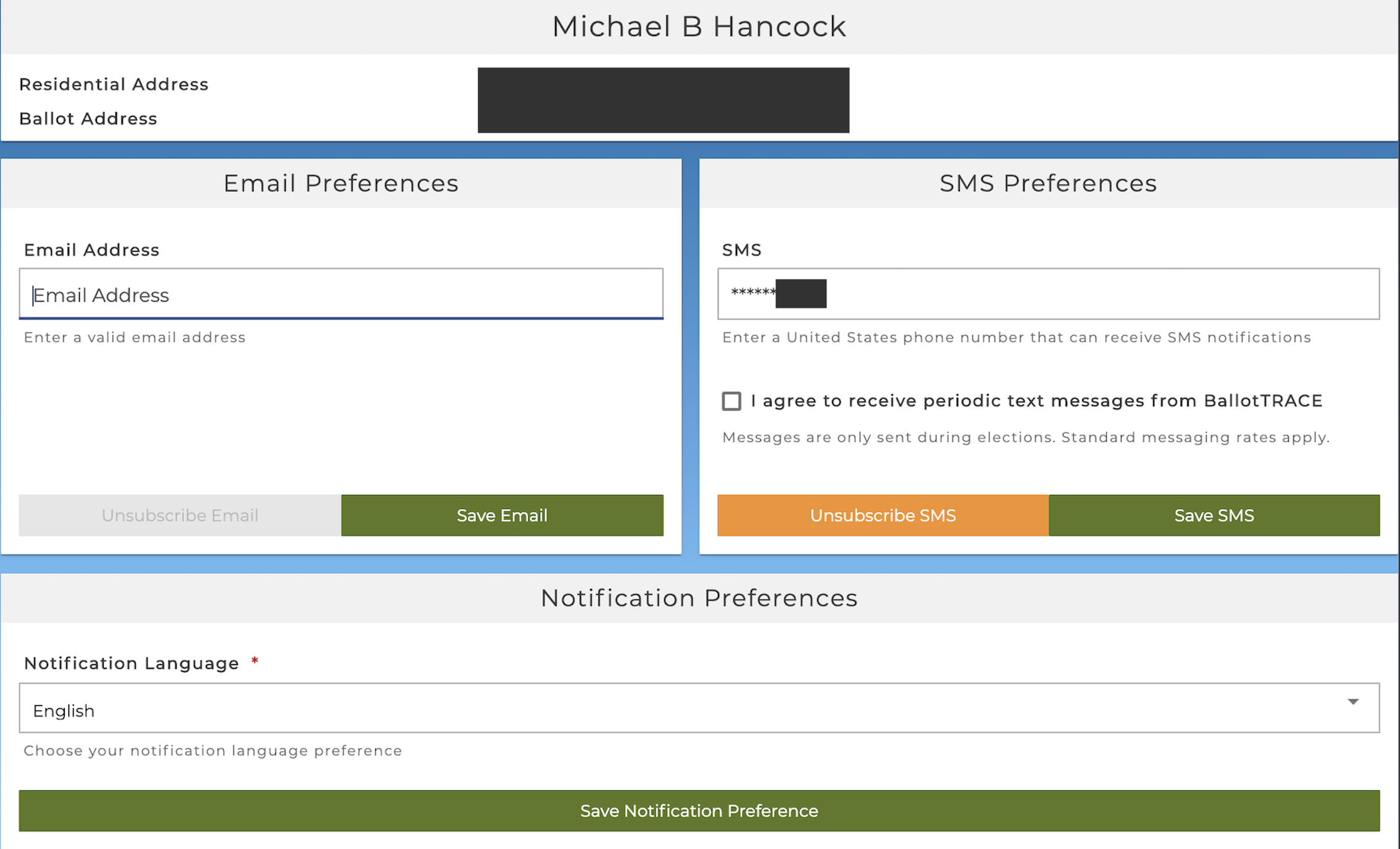}\\
    \caption{BallotTRACE's notifications page for the mayor of Denver, CO.  The system appears to allow any user who accesses a voter information page to update voter notifications.  Again, sensitive information is redacted.}
\end{figure*}

BallotTRACE, developed by i3logix in Denver, Colorado, operates similarly to BallotTrax.  Figure 1 displays the lookup form used by BallotTRACE and shows that the system allows voter lookup based only on first name, last name, ZIP code, and birth year \cite{ballotTRACE}.  To demonstrate the feasibility of arbitrary voter lookup, we chose to look up the voter information of the Mayor of Denver, Michael Hancock, as an example.

Using voter record data from VoterRecords.com, we were able to access the mayor's ballot tracking status account page, and further view the voting history for the 2020 primary, as shown in Figure 3 \cite{ballotTRACE}.  We were further able to access the “Notification Preferences” page, as shown in Figure 4, and seemingly could have modified or unsubscribed from notification updates.  Even without making any modifications to contact information, an adversary could view a voter's email address or partial phone number, as demonstrated.

The problem is fundamentally one of insufficient authentication.  The system cannot guarantee with any measure of confidence that the user looking up a particular voter’s information is truly the voter if the only information required to look up a voter is publicly available.  

Our proposed solution is to recommend the use of a 12-digit unique, randomized ID that is assigned to a particular ballot envelope.  The concept is similar to the United States’ 2020 Census use of 12-digit Census IDs, which are included in the materials mailed to residents \cite{census}.  A state or county can provide a voter with this unique ID as part of their paper-based or online absentee ballot request.  For states such as California that are planning on automatically sending all voters mail-in ballots, this ID could be included with the ballot to enable voters to track its return to their local election facility \cite{caliBallot}.  Voters could use this randomly assigned ID along with typical identifying information, such as first name and last name, to authenticate themselves to the ballot tracking system.

\subsection{Web Lookup Form Security}
Like online voter registration systems, the various ballot tracking web applications all necessarily contain HTML or AngularJS forms in which voters can enter their information.  This input is then used as the basis for subsequent SQL queries to the election facility’s ballot tracking database \cite{caliBallotTrax, ballotTRACE, ballotScoutLookup}.  Given this proximity to an important elections database, properly securing these forms is critical.

For a broad overview of the configuration of each platform’s TLS/SSL web server, we used Qualys’ public SSL Server Test scanning tool, which evaluates a website on the basis of its certificate, protocol support, key exchange, and cipher strength \cite{sslServerTest}.

\textbf{BallotTRACE} BallotTRACE’s certificate signature algorithm uses SHA256 with RSA-4096.  It only supports TLS 1.2 and above and is therefore secure against attacks affecting older versions of SSL/TLS like DROWN or POODLE \cite{sslServerTest}.

\textbf{BallotTrax}  California’s BallotTrax system uses a signature algorithm with SHA256 and RSA with a 2048-bit key.  However, the server supports TLS 1.1, a legacy version of TLS which has been shown to be insecure \cite{sslServerTest}.  Chrome is planning on deprecating support for TLS 1.0 and 1.1 in 2020, citing flaws in MD5 and SHA-1, both used by these older versions of TLS.  Apple, Microsoft, and Mozilla announced similar plans \cite{oldTLS}.

\textbf{Ballot Scout}  Ballot Scout’s signature algorithm also uses SHA256 and RSA with a 2048-bit key.  Ballot Scout supports protocols TLS 1.2 and 1.3 and prevents downgrade and other common attacks \cite{sslServerTest}.

Overall, BallotTRACE and Ballot Scout's server configurations provide basic levels of security, although BallotTrax’s configuration is flawed due to its support of a legacy version of TLS. Scanning all three sites for SQL injection vulnerabilities using Pentest-Tools’ SQL Injection Scanner in addition to manual input testing revealed no SQL injection vulnerabilities \cite{sqliTest}.  While there are no automated scanners capable of detecting all possible vulnerabilities, these results are a promising indicator of solid security practices.

\subsection{Barcode Security}
As previously mentioned when discussing USPS services, mail-in ballot tracking is enabled through the use of Intelligent Mail barcodes (IMBs), developed by the United States Postal Service.  Mailed ballots are contained in an outer envelope with a 65-bar Intelligent Mail barcode, the technical specifications of which are publicly available \cite{imbTechnical}.  When a completed ballot is mailed back, a machine at a central elections facility scans the barcode and updates the ballot tracking information accordingly.  

A barcode is fundamentally an input to a system.  As such, barcodes represent a potential vulnerability.  While an effective security policy naturally distrusts system inputs and assumes the possibility of a malicious adversary, barcode scanners have not historically adopted this attitude of distrust.  A 2008 talk at DEFCON demonstrated the feasibility of multiple barcode-driven attacks, including barcode-driven buffer overflow, SQL injection, and cross-site scripting attacks, and specifically mentioned Intelligent Mail barcodes as an example \cite{toyingBarcodes}. QR codes were also shown in 2012 to be capable of exploiting vulnerabilities in the reader software or operating system, such as SQL injections \cite{qrCodes}. Since then, the capabilities and motivations of malicious adversaries have only increased, but there has been no evidence that barcode security has seen a commensurate increase in attention.

A barcode, then, is actually an attack vector into a system.  A maliciously crafted barcode is capable of launching common security attacks.  In the ballot tracking process, the concern is that an adversary could create a malicious barcode and mail back an envelope with this barcode instead of a genuine Intelligent Mail barcode.  This adversarial barcode would then be scanned at an elections facility, conceivably causing damage to election databases tracking which ballots have been returned---and, far more consequentially, who has already voted in the election.  To mitigate potential consequences of a malicious barcode, then, the application that takes in scanner data should validate and sanitize all inputs, whether in text or barcode form.

\section{Information Privacy}
In order to properly deliver results of ballot tracking to a voter, ballot tracking services often require the submission of personal data. In order to deliver updates on a voter's ballot, Ballot Scout requires the submission of a voter's name, address, year of birth, and email address or phone number. Ballot Scout notes in their privacy policy that they "do not disclose any of your personal information unless required by law" \cite{ballotScout}.

While in many states voter information is public, the amount of data varies between states and the means by which to obtain the information are different as well \cite{voterLists}. This is where a crucial distinction lies. It might be very possible to obtain voter information in many states, but there is often a process by which one must request a list of voter information. As a result, there is a paper trail that could pin responsibility. While that does not directly protect privacy in and of itself, the fact that any malicious act as a result of this information request  could be traced back to the requester is thwarting. When methods are introduced by which to obtain this information in an anonymous way via hacking, the privacy of the voting public is put at greater risk. 

The schema and means by which the information is stored is not publicly available, as these ballot tracking services are mostly powered by private corporations. Presumably, this is intended to ensure the confidentiality of proprietary designs.  However, this does pose the privacy concern of aggregate voter statistics. As it would be necessary to track one’s ballot, the ballot tracking services are made aware of distinct voter actions. These actions can be defined as but are not limited to: whether or not the individual has voted, at what time they voted, where they chose to deposit their vote. While the contents of the ballot might remain secure, the behaviors of voters may not. Similar to Ballot Scout, many tracking services explicitly state that they will not give away your personal information. However, there are no such lines in the privacy policies describing the usage or sale  of aggregate statistics obtained by tracking the voter ballots \cite{ballotScout}.  

This brings up the issue of differential privacy: wherein, the aggregate statistics must be able to describe patterns of groups within the dataset while withholding personally identifiable information, in this case the actions of the individual. If the company were to sell aggregate information that could isolate individuals, it becomes much easier for political entities to interfere with the mail-in voting process. This becomes vulnerable to an attack such as a Membership Attack in which the attacker can determine whether or not a specific individual falls within a subset of the data. For example, an adversary would be able to determine, via membership attack,  a set of individuals that do not vote by mail. In a state where vote by mail is the only option, contextual information such as this example becomes very powerful in determining the behavior of voters. Similarly, since party affiliation is publicly known, with the voter actions known of a subset of individuals, the following scenario could occur: Time or location data for mail-in ballots are released or sold to Party A. Party A, via membership attack, isolates and learns that Party B voters tend to deposit their ballots at a certain time or location more frequently. As a result, Party A tampers with deposit boxes in that location for only those specific times to achieve a maximal impact result with minimal effort.  
 
The problem of privacy becomes compounded when the security of said information is subjected to multiple weak points. Ballot tracking services are not powered entirely by the technology of the given corporation. As a result they are not solely responsible for all of the potential security breaches. Trackers require the usage of third party tools in order to deliver their end product. In the case of Ballot Scout, to deliver its tracking services it enlists the services of the following services: Twilio, Amazon SES, and SmartyStreets \cite{ballotScout}. Each of these distinct services gain access to different pieces of a voter's personal information. These services similarly rely on other 3rd party services. As a result, a chain of dependencies is created in which a voter’s personal data is exposed at multiple different points. 

Twilio, for example, had a breach in 2018 from one of its providers, Voxox, that resulted in SMS message details being leaked \cite{twilioBreach}. Since Twilio is used to power voter notifications, a hack on any part of this pipeline could result in the adversary being aware of where the ballot is due to these notifications. The adversary could similarly obtain authentication codes to register ballot tracking properly. It seems imperative that there must be accountability for the pipeline of providers that have access to voter’s personal information. Otherwise, the public must be adequately educated on the reach that their personal information has when being provided to a ballot tracking service such as Ballot Scout. 

\section{Conclusion}

Vote-by-mail has received increased scrutiny during the COVID-19 pandemic, with widespread implementation seen as a necessity given social distancing restrictions. States that have implemented no-excuse, state-wide vote-by-mail have seen vast successes and with higher voter turnout, highlighting its efficacy as an alternative to traditional on-site ballot casting. 

Security concerns and lack of transparency with the chain of custody are often cited as the primary reason for sticking to traditional voting methods. Voters want to ascertain their ballots are actually counted, and for good reason as on-site voting machines already have numerous security vulnerabilities \cite{electionMachines}. Attempts to rectify this have produced numerous web and mobile applications, including Ballot Scout, BallotTRACE, and BallotTrax, that provide users with an interface to view ``end-to-end mail tracking information" \cite{uspsIVsite}. Unfortunately, any technological augmentation to a paper-based voting scheme is itself a potential security vulnerability. 

In performing security analyses on these applications and other relevant sites, we have reaffirmed that although there exist concerns—--including the use of weak user authentication and online voter registration site schemes—--none present vulnerabilities that can be exploited on a large scale to directly influence an election. The example fraudulent account we temporarily created for USPS, for instance, requires an attacker to leave an information trail with their respective phone carrier, and is not scalable.  Although ballot tracking systems pose concerning privacy questions, it would be difficult for an adversary to use them to perpetrate voter fraud on a large scale.

Our findings entirely support the notion that vote-by-mail is an ideal scheme for wide-spread implementation, despite its flaws related to tracking. Vote-by-mail offers a robust paper-trail and has been shown to increase voter turnout and engagement \cite{vahGuide}. The electronic systems supporting remote voter registration and voting, however, will need significant security improvements before we can truly trust them to uphold the integrity of our democratic processes.

\section{Acknowledgements}
We want to express our appreciation to the 6.857 staff for their dedication to instruction and support in what will invariably be a semester to remember.

We would like to especially thank Ron Rivest for his helpful discussions, guidance, and encouragement. We are also grateful for the volunteers local to states using these ballot tracking systems for assisting us in our evaluation, and the volunteer that allowed us to sign up for Informed Delivery using their home address.

\printbibliography

@online{pewTrustsUnderstanding,
    title = {Understanding Online Voter Registration},
    author = {The Pew Charitable Trusts},
    year = {2013},
    url = {https://www.pewtrusts.org/~/media/legacy/uploadedfiles/pcs_assets/2013/UnderstandingOnlineVoterRegistrationpdf.pdf},
}

@online{rockTheVote,
    title = {2018 Annual Report},
    author = {Rock the Vote},
    year = {2018},
    url = {https://www.rockthevote.org/wp-content/uploads/Rock-the-Vote-2018-Annual-Report.pdf},
}

@online{pewTrustsOVR,
    title = {Online Voter Registration: Trends in development and implementation},
    author = {The Pew Charitable Trusts},
    year = {2015},
    url = {http://www.pewtrusts.org/~/media/Assets/2015/05/OVR_2015_brief.pdf?la=en},
}

@online{uniqueID,
    title = {Unique ID},
    author = {Alan De Smet},
    year = {2013},
    url = {http://www.highprogrammer.com/cgi-bin/uniqueid/dl_md},
}

@online{mdRegistration,
    title = {Voter Registration},
    author = {Maryland State Board of Elections},
    year = {2020},
    url = {https://elections.maryland.gov/voter_registration/index.html},
}

@online{harvardPaper,
    title = {Voter Identity Theft: Submitting Changes to Voter Registrations Online to Disrupt Elections},
    author = {Latanya Sweeney and Ji Su Yoo and Jinyan Zang},
    year = {2017},
    organization = {Technology Science},
    url = {https://techscience.org/a/2017090601},
}

@online{ncslSecuring,
    title = {Securing Voter Registration Systems},
    author = {Dylan Lynch},
    year = {2018},
    organization = {National Conference of State Legislatures},
    url = {https://www.ncsl.org/research/elections-and-campaigns/securing-voter-registration-systems.aspx},
}

@online{primaryHack,
    title = {DA: Hackers Penetrated Voter Registrations in 2016 Through State's Election Site},
    author = {John Sepulvado},
    publisher = {KQED},
    year = {2017},
    url = {https://www.kqed.org/news/11579541/hackers-penetrated-voter-registrations-in-2016-through-states-election-site},
}

@online{ncslInterview,
    title = {Interview with J. Alex Halderman on Cybersecurity for Online Voter Registration},
    author = {National Conference of State Legislatures},
    year = {2013},
    url = {https://www.ncsl.org/research/elections-and-campaigns/itnerview-j-alex-halderman-online-registration.aspx},
}

@online{nist,
    title = {Security Best Practices for the Electronic Transmission of Election Materials for UOCAVA Voters},
    author = {National Institute of Standards and Technology},
    year = {2011},
    organization = {U.S. Department of Commerce},
    url = {https://www.nist.gov/system/files/documents/itl/vote/nistir7711-Sept2011.pdf},
}

@online{zombiePOODLE,
    title = {New Zombie 'POODLE' Attack Bred from TLS Flaw},
    author = {Kelly Jackson Higgins},
    year = {2019},
    url = {https://www.darkreading.com/vulnerabilities---threats/new-zombie-poodle-attack-bred-from-tls-flaw/d/d-id/1333815?_mc=sm_iwfs_editor_kellysheridan},
}

@conference{weakDH,
    title = {Imperfect Forward Secrecy: How Diffie-Hellman Fails in Practice},
    author = {David Adrian and Karthikeyan Bhargavan and Zakir Durumeric and Pierrick Gaudry and Matthew Green and J. Alex Halderman and Nadia Heninger and Drew Springall and Emmanuel Thom\'{e} and Luke Valenta and Benjamin VanderSloot and Eric Wustrow and Santiago Zanella-B\'{e}guelin and Paul Zimmermann },
    year = {2015},
    publisher={Proceedings of the 22nd ACM SIGSAC Conference on Computer and Communications Security}
}

@online{forwardSecrecy,
    title = "SSL and TLS Deployment Best Practices",
    author = "SSL Labs",
    year = {2020},
    url = {https://github.com/ssllabs/research/wiki/SSL-and-TLS-Deployment-Best-Practices},
}

@online{sslServerTest,
    title = {SSL Server Test},
    author = {SSL Labs},
    year = {2020},
    organization = {Qualys, Inc.},
    url = {https://www.ssllabs.com/ssltest/},
}

@online{uspsAudit,
    title = {Informed Visibility Vulnerability Assessment},
    author = {Office of Inspector General},
    year = {2018},
    organization = {United States Postal Service},
    url = {https://www.uspsoig.gov/sites/default/files/document-library-files/2018/IT-AR-19-001.pdf},
}

@online{krebsIVMTR,
    author = {Brian Krebs},
    title = {USPS Site Exposed Data on 60 Million Users},
    url = {https://krebsonsecurity.com/2018/11/usps-site-exposed-data-on-60-million-users/},
    year = {2018},
}

@online{krebsID2017,
    author = {Brian Krebs},
    title = {USPS `Informed Delivery' Is Stalker's Dream},
    url = {https://krebsonsecurity.com/2017/10/usps-informed-delivery-is-stalkers-dream/},
    year = {2017},
}

@online{krebsID2018,
    author = {Brian Krebs},
    url = {https://krebsonsecurity.com/2018/11/u-s-secret-service-warns-id-thieves-are-abusing-uspss-mail-scanning-service/},
    title = {US Secret Service Warns ID Theieves are Abusing USPS's Mail Scanning Service},
    year = {2018},
}

@online{uspsSignup,
    title = {How to Sign Up for Informed Delivery},
    url = {https://www.usps.com/c360/images/informed_delivery/Informed\%20Delivery\%20Sign\%20Up\%20Guide\%20Jan\%202020.pdf},
    author = {United States Postal Service},
    year = {2020},
    }

@online{uspsMobilePolicy,
    title = {Online Mobile Phone Verification},
    url = {https://ips.usps.com/IPSWeb/verification_user_information.xhtml}}

@online{aclu,
    url = {https://www.aclu.org/report/aclu-report-voting-rights-act},
    pages = {34-53},
    author = {American Civil Liberties Union},
    publisher = {American Civil Liberties Union},
    title = {The Case for Restoring and Updating the Voting Rights Act: A Report of the American Civil Liberties Union.}
    }

@online{electionMachines,
    url = {https://www.brennancenter.org/our-work/analysis-opinion/voting-machine-security-where-we-stand-six-months-new-hampshire-primary},
    title = {Voting Machine Security: Where We Stand Six Months Before the New Hampshire Primary},
    year = {2019},
    author = {Andrea Cordova McCadney and Elizabeth Howard and Lawrence Norden}}

@online{ballotScout,
    author = {Democracy Works},
    title = {Ballot Scout Privacy Policy},
    url = {https://www.democracy.works/ballot-scout-privacy-policy-terms-of-service},
    }

@online{uspsIVsite,
    author = {United States Postal Service},
    title = {Informed Visibility Mail Tracking \& Reporting},
    url = {https://iv.usps.com/#/landing},
    year = {2020}}

@online{usps2018APIcopy,
    author = {United States Postal Service},
    title = {2018 Informed Delivery API},
    url = {https://krebsonsecurity.com/wp-content/uploads/2018/11/USPS-ID-API.txt},
    year = {2018}}

@online{usps2019APIcopy,
    author = {United States Postal Service},
    title = {2019 Informed Delivery API Documentation},
    url = {https://mailomg.files.wordpress.com/2019/08/iv-mtr-api-developer-toolkit_v2.5.pdf},
    year = {2019}}

@online{NARF,
    author = {NARF},
    title = {North Dakota Agrees to Court-ordered Relief Easing Voter ID Laws for Native Americans on Reserations},
    year = {2020},
    url = {https://www.narf.org/nd-voting-rights/},
    publisher = {Native American Rights Fund}
    }

@online{scaleVAH,
    publisher = {National Vote at Home Institute},
    title = {Vote at Home Scale Plan},
    year = {2020},
    author = {Natonal Vote at Home Institute},
    url = {https://www.voteathome.org/wp-content/uploads/2020/03/VAHScale_StrategyPlan.pdf},
    }

@online{twilioBreach,
    author={Twilio},
    title={Twilio response to Voxox data breach},
    url={https://www.twilio.com/blog/twilio-response-to-voxox}
}

@online{voterLists,
    author={National Conference of State Legislatures},
    title={Access To and Use of Registration Voter Lists},
    url={https://www.ncsl.org/research/elections-and-campaigns/access-to-and-use-of-voter-registration-lists.aspx}}

@online{VBM,
    author={The New York Times Editorial Board},
    organization={The New York Times},
    title={The 2020 Election Won't Look Like Any We've Seen Before},
    url={https://www.nytimes.com/2020/03/21/opinion/sunday/coronavirus-vote-mail.html?referringSource=articleShare}}

@online{WydenKlobuchar,
    author={Amy Klobuchar and Ron Wyden},
    organization={Washington Post},
    title={Here's how to guarantee coronavirus won't disrupt our elections},
    url={https://www.washingtonpost.com/opinions/2020/03/16/heres-how-guarantee-coronavirus-wont-disrupt-our-elections/}}

@online{internetVoting,
    url = {https://people.csail.mit.edu/rivest/pubs/PSNR20.pdf},
    title = {Going from Bad to Worse: From Internet Voting to Blockchain Voting},
    year = {2020},
    author = {Sunoo Park and Michael Specter and Neha Narula and Ronald L. Rivest}}

@conference{voatz,
    title={The Ballot is Busted Before the Blockchain: A Security Analysis of Voatz, the First Internet Voting Application Used in U.S. Federal Elections},
    author={Michael A. Specter and James Koppel and Daniel Weitzner},
    year={2020},
    publisher={Proceedings of the 29th USENIX Security Symposium}}

@online{democracyWorks,
    organization={Democracy Works},
    author={Democracy Works},
    url={https://www.democracy.works/},
}

@online{ballotTRACEProgram,
    author={Denver Elections Division},
    title={2010 Professional Practices Program: Ballot TRACE},
    url={https://www.electioncenter.org/publications/2010\%20PPP/Denver_Election\%20Paper\%20Submittal_Ballot\%20Trace_2010.pdf}}

@online{fastCompany,
    author={Talib Visram},
    organization={Fast Company},
    title={Track your ballot like a package: How technology will smooth the way for November's mail-in ballot surge},
    url={https://www.fastcompany.com/90501588/track-your-ballot-like-a-package-how-technology-will-smooth-the-way-for-novembers-mail-in-ballot-surge}}

@conference{nationalAcademies,
    publisher={The National Academies Press},
    author={National Academies of Science and Engineering and Medicine},
    title={Securing the Vote: Protecting American Democracy},
    date=2018,
    month=sep,
    day=6}

@online{guardian,
    author={Kim Zetter},
    organization={The Guardian},
    title={US government plans to urge states to resist 'high-risk' internet voting},
    url={https://www.theguardian.com/us-news/2020/may/08/us-government-internet-voting-department-of-homeland-security},
    date=2020,
    month=may,
    day=8}

@online{nycVoter,
    author={Vivian Wang},
    organization={The New York Times},
    title={Public Records: Personal Information on New York City Voters is Now Available for All to See},
    url={https://www.nytimes.com/2019/04/26/nyregion/voter-registration-nyc-online.html},
    date=2019,
    mon=apr,
    day=26}

@online{nycDatabase,
    author={Stephen P. Morse},
    title={Searching the New York State Voter Records in One Step (2002-2019)},
    url={https://stevemorse.org/nysvoters/nysvoters.html}}

@online{caliBallotTrax,
    organization={BallotTrax},
    title={Where's My Ballot?},
    url={https://california.ballottrax.net/voter/}}

@online{ballotTRACE,
    organization={City and County of Denver},
    title={BallotTRACE:Tracking, Reporting And Communication Engine},
    url={https://ballottrace.org/home}}

@online{ballotScoutLookup,
    organization={Democracy Works},
    title={Ballot Scout Lookup Widget},
    url={https://www.democracy.works/ballot-scout-lookup-widget-sample}}

@online{voterRecords,
    organization={VoterRecords.com},
    title={Voter Registration Records},
    url={https://voterrecords.com/}}

@online{census,
    organization={United States Census 2020},
    url={https://my2020census.gov/login}}

@online{imbTechnical,
    organization={United States Postal Service},
    title={Intelligent Mail Barcode Technical Resource Guide},
    url={https://postalpro.usps.com/node/221},
    date=2009,
    mon=jan,
    day=13}

@online{toyingBarcodes,
    organization={DEFCON},
    title={Toying with Barcodes},
    url={https://www.youtube.com/watch?v=qT_gwl1drhc},
    date=2011}

@online{sqliTest,
    organization={Pentest-Tools.com},
    title={SQL Injection Scanner},
    url={https://pentest-tools.com/website-vulnerability-scanning/sql-injection-scanner-online#}}

@online{voteShaming,
    organization={The New York Times},
    title={Did You Vote? Now Your Friends May Know (and Nag You)},
    author={Natasha Singer},
    url={https://www.nytimes.com/2018/11/04/us/politics/apps-public-voting-record.html},
    year=2018,
    mon=nov,
    day=4}

@online{stateShaming,
    organization={The Washington Post},
    title={Registered to vote?  Your state may be posting personal information about you online},
    author={Aki Peritz},
    url={https://www.washingtonpost.com/outlook/2019/04/09/registered-vote-your-state-is-posting-personal-information-about-you-online/},
    year=2019,
    mon=apr,
    day=9}

@online{caliBallot,
    organization={The New York Times},
    title={California to Mail All Voters Ballots for November Election},
    author={Nick Corasaniti and Jennifer Medina},
    url={https://www.nytimes.com/2020/05/08/us/politics/california-mail-vote-november-election.html},
    year={2020}}

@online{oldTLS,
    organization={Google Security Blog},
    title={Modernizing Transport Security},
    author={David Benjamin},
    url={https://security.googleblog.com/2018/10/modernizing-transport-security.html},
    year={2018}}

@online{vahGuide,
    organization={National Vote at Home Institute (NVAHI)},
    title={Vote at Home Policy and Research Guide},
    url={https://www.voteathome.org/wp-content/uploads/2019/03/VAH-Policy-and-Research-Guide.pdf},
    year=2020}

@online{qrCodes,
    title={Malicious Pixels Using QR Codes as Attack Vector},
    author={Peter Kieseberg and Sebastian Schrittwieser and Manuel Leithner and Martin Mulazzani and Edgar Weippl and Lindsay Munroe and Mayank Sinha},
    organization={Trustworthy Ubiquitous Computing pp. 21-28},
    url={https://link.springer.com/chapter/10.2991/978-94-91216-71-8_2},
    year={2012}}

\end{document}